\documentclass[10pt,a4paper]{article}

\usepackage{times}

\usepackage{epsfig,wrapfig}
\usepackage{epstopdf} 

\usepackage{fancyhdr}
\usepackage{color}

\usepackage[lmargin=2.5cm, rmargin=2.5cm,tmargin=2.50cm,bmargin=2.50cm]{geometry}
\usepackage{indentfirst}

\usepackage{titlesec}
\titleformat{\section}[hang]
  {\centering}{\thesection}{1ex}{\normalsize \textsc}
\titleformat{\subsection}[hang]
  {}{\thesubsection}{1ex}{\normalsize \textit}

\newcommand{\acknowledgement}{\section*{\centering{\textnormal{\normalsize{\textsc{Acknowledgement}}}}}}

\renewcommand{\thesection}{ \normalsize \textnormal{\Roman{section}.}}
\renewcommand{\thesubsection}{\normalsize \textnormal{\textsc{\textit{\Alph{subsection}.}}}}

\def\e{\begin{equation}}
\def\f{\end{equation}}
\def\_#1{{\bf #1}}

\def\.{\cdot}

\def\VCC{V_{\mathrm{CC}}}
\def\VEE{V_{\mathrm{EE}}}
\def\Cpn{C_{\mathrm{p},n}}
\def\Vo{V_{\mathrm{o}}}
\def\RLn{R_{\mathrm{L},n}}
\def\dd{\mathrm{d}}
\def\Cp{C_\mathrm{p}}
\def\RL{R_\mathrm{L}}
\def\Vr{V_\mathrm{r}}

\begin{document}
\title{\large \textbf{Active Metasurfaces as a Platform for Capacitive Wireless Power Transfer Supporting Multiple Receivers}}
\def\affil#1{\begin{itemize} \item[] #1 \end{itemize}}
\author{\normalsize \bfseries \underline{Fu Liu}, Prasad Jayathurathnage and Sergei A. Tretyakov
}
\date{}

\maketitle
\thispagestyle{fancy} 
\vspace{-6ex}
\affil{\begin{center}\normalsize Department of Electronics and Nanoengineering, Aalto University, P.O. Box 15500, Espoo, Finland \\
{\color{blue} fu.liu@aalto.fi}
 \end{center}}

\begin{abstract}
\noindent \normalsize
\textbf{\textit{Abstract} \ \ -- \ \
As wireless power transfer (WPT) repeaters, metasurfaces can enhance field coupling, improving the WPT operation. In this paper, we show that metasurfaces can also be used as transmitters of capacitive WPT systems. Such a metasurface-based WPT system can feed  multiple receivers and provide robust operation against load or position variations. We formulate an analytical model of such WPT systems. We also discuss the exact solution of a particular example with $N$ identical receivers.
}
\end{abstract}

\section{Introduction}
Integrated with lumped elements or mixed-signal integrated circuits, tunable metasurfaces have shown great capability to manipulate electromagnetic waves and a vast spectrum of applications have been achieved, such as polarization conversion, wavefront shaping, holography, and tunable perfect anomalous reflections~\cite{TunableMetasurface,itelligentMetasurface}. On the other hand, wireless power transfer (WPT) is an emerging and fast growing research topic and metasurfaces (and metamaterials) can be introduced to WPT systems to enhance the system efficiency by engineering local evanescent fields~\cite{metamaterialWPT}. For example, a wire medium slab can convert evanescent waves to propagating modes and therefore it can be used to increase magnetic coupling between two distant coils. Based on this idea, a smart table, which can support multiple receivers, has been constructed by using a planar version of wire media~\cite{smartTable}. In this paper, rather than using magnetic coupling, we show that a tunable metasurface can work as a smart table which can wirelessly transfer power to multiple receivers with capacitive coupling. Moreover, the proposed solution ensures  robust WPT operations subjected to changes of receiver position and load resistance.

\begin{figure}[b!]
	\centering 
	\includegraphics[width=0.55\textwidth]{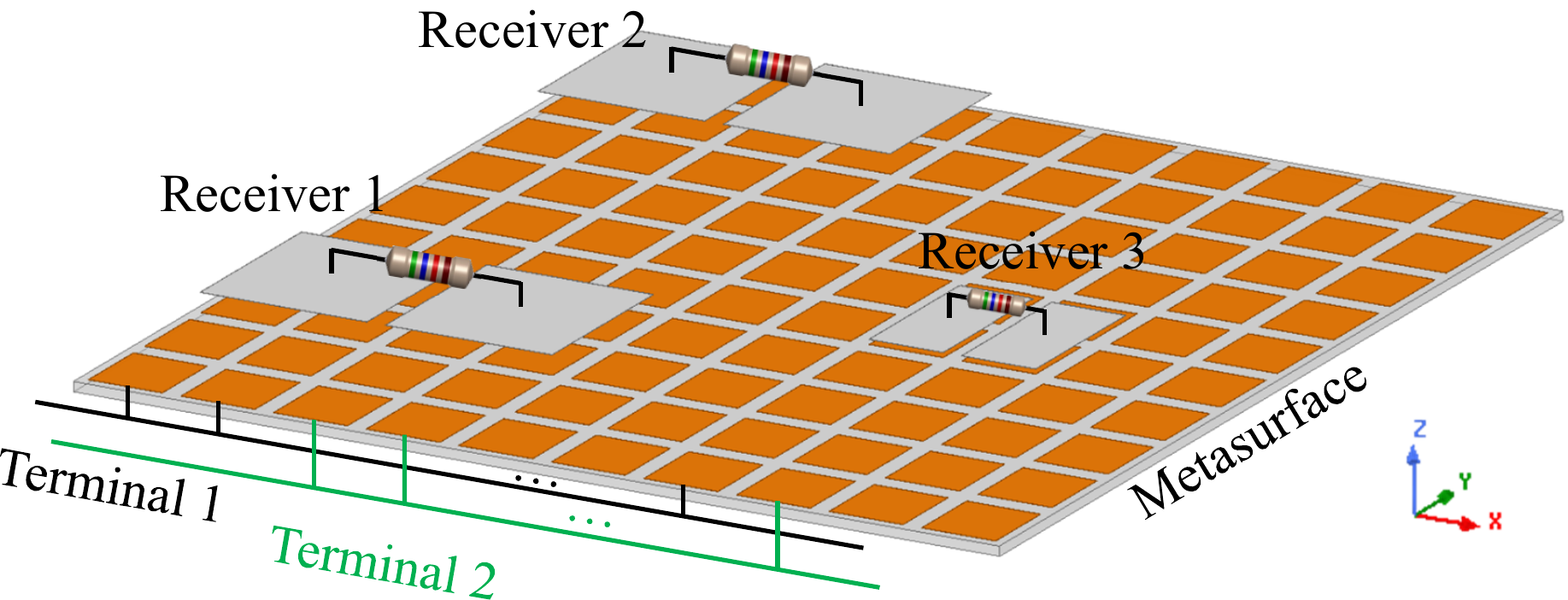}
	\caption{Schematic of the metasurface-based WPT system supporting multiple receivers. 
	}\label{Fig1}
\end{figure}

Recent studies have shown that robust WPT operations can be achieved by using a capacitive self-adaptive WPT system, in which the operational amplifier (op-amp) works as a switch. Robustness is ensured by a special circuit topology where the coupling capacitors and the load are parts of the feedback loop~\cite{onsiteWPT,selfadaptiveWPT}. Therefore, when the receiver position (equivalently, the capacitive coupling strength) or load resistance change, the system can automatically adjust itself to the optimal working condition. Metasurfaces formed by  patch arrays can be naturally used to form capacitors, thus, they can be used as transmitters for a capacitive WPT system. Moreover, many patches of the metasurface can support multiple receivers. Such a working arrangement is schematically shown in Fig.~\ref{Fig1}, where different rows of patches on the metasurface are connected to terminal~1 (the output terminal of op-amp) or terminal~2 (the inverting terminal of the op-amp) of a self-adaptive WPT system, see Fig.~\ref{Fig2}(a). Whenever a receiver is present and the self-oscillating condition is satisfied, a feedback loop through that receiver is  formed and  power is wirelessly transferred to the load through the capacitive link. 

\section{Analytical Model of the Metasurface Based WPT System}
\begin{figure}[t!]
	\centering 
	\includegraphics[width=\textwidth]{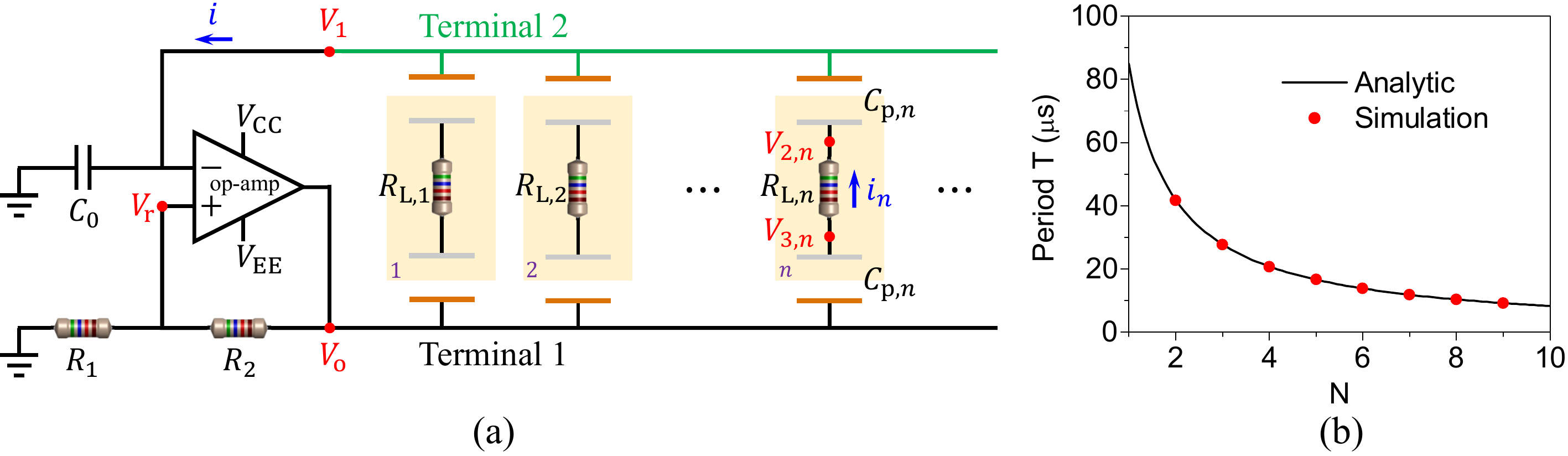}
	\vspace{-7mm}
	\caption{(a) Circuit diagram of the metasurface-based capacitive WPT system utilizing the self-oscillating approach. The notations of the voltages and currents are also shown. (b) The oscillation period for different numbers of identical receivers. The solid line (symbols) are from the analytical formula Eq.~\ref{Eq:T} (LTSpice simulation).}
	\label{Fig2}
\end{figure}

The circuit diagram of the metasurface-based capacitive WPT system supporting multiple receivers is shown in Fig.~\ref{Fig2}(a). For simplicity, we assume that the op-amp is ideal, i.e., lossless with infinite slew rate and infinite input impedance. We also assume that the coupling capacitance of the two capacitors for each receiver, formed by the patches on the metasurface and the patches on the receiver, are the same and denote them as $\Cpn$ for the $n^{\mathrm{th}}$ receiver, as shown in Fig.~\ref{Fig2}(a). When the system operates, capacitors $C_0$ and $\Cpn (n=1,2,...)$ undergo charging and discharging processes due to non-zero and different voltages $\Vo$ and $V_1$. By comparing $V_1$ to the reference voltages $\Vr$, the op-amp will switch the output $\Vo$ between $\VCC$ and $\VEE$ ($=-\VCC$ in this work for symmetric operation). Therefore voltage oscillations are formed and power is  wirelessly generated at each load position~\cite{selfadaptiveWPT}.

In such a multiple receiver WPT system, the master equations can be formulated from the circuit theory. First of all, for each receiver, the voltage drops across the two coupling capacitors are equal because the capacitances are equal, i.e., $\Vo-V_{3,n}=V_{2,n}-V_1$, which gives $V_{2,n}=\Vo+V_1-V_{3,n}$. Secondly, as the current $i_n$ through the two coupling capacitors and the load $\RLn$ are identical, we obtain
\begin{equation}
	i_n=\Cpn \frac{\dd (\Vo-V_{3,n})}{\dd t}=\frac{2V_{3,n}-\Vo-V_1}{\RLn},\ \ \ \ \ \  (n=1,2,...,N)
	\label{eq:master1}
\end{equation}
for each receiver, where $N$ is the total number of receivers. Finally, from  Kirchhoff's current law, the total current through the capacitor $C_0$ is the sum of the currents through all the $N$ receivers (note that zero current flows into the op-amp due to its  infinite input impedance), giving 
\begin{equation}
	C_0 \frac{\dd V_1}{\dd t}=\sum_{n=1}^{N} \frac{2V_{3,n}-\Vo-V_1}{\RLn}.
	\label{eq:master2}
\end{equation} 
Equations (\ref{eq:master1}) and (\ref{eq:master2}), in total $N+1$ equations, are the master equations of the WPT system with multiple receivers shown in Fig.~\ref{Fig2}(a). For a particular configuration, $C_0, \Cpn, \RLn, \Vo$, and reference voltage $\Vr=\beta\Vo$, where $\beta=R_1/(R_1+R_2)$ is given by the voltage divider resistors, are known and there are $N+1$ unknown variables, i.e., $V_1$ and $V_{3,n}$. Therefore, the system can be solved with proper initial conditions. In the next section, we give one example for many identical receivers, providing an  analytical solution.

\section{Exact Solution for $N$ Identical Receivers}
When all $N$ receivers are identical, i.e., same coupling $\Cp$ and load $\RL$, the total current $i$ will be equally distributed to the loads, and the master equations (\ref{eq:master1}) and (\ref{eq:master2}) can be simplified into
\begin{eqnarray}
	-\Cp \frac{\dd V_3}{\dd t} & = & \frac{2V_3-\Vo-V_1}{\RL},\\
	C_0 \frac{\dd V_1}{\dd t} & = & N\frac{2V_3-\Vo-V_1}{\RL}.
\end{eqnarray}
On the other hand, the initial conditions can be set when the output voltage switches from $\VCC$ to $\VEE$, which gives $\Vo|_{t=0}=-\VCC$, $V_1|_{t=0}=\beta \VCC$, and $V_3|_{t=0}=-(1+\beta C_0/(N\Cp))\VCC$. Therefore, the solution for this WPT system with $N$ identical receivers is analytically obtained as
\begin{eqnarray}
	V_1(t) & = & \frac{\VCC}{2 C_0+N\Cp} \left((N\Cp+2\beta
		C_0+N\beta\Cp)e^{-\frac{2C_0+N\Cp}{C_0\Cp\RL}t}-N\Cp\right),\\
	V_3(t) & = & -C_0/(N\Cp)V_1(t)-\VCC.
\end{eqnarray}
Similarly to~\cite{selfadaptiveWPT}, we can find the oscillation period (from $V_1(T/2)=-\beta\VCC$), the oscillation condition (from $V_1(t\rightarrow\infty)<-\beta\VCC$), and the averaged transferred power to each load (from $1/T\int_{0}^{T} (V_3-V_2)^2/\RL \dd t$) as
\begin{eqnarray}
	T & = & 2\frac{C_0 \Cp \RL}{2C_0+N\Cp}	\ln\left[\frac{N\Cp+(2C_0+N\Cp)\beta}{N\Cp-(2C_0+N\Cp)\beta}\right],\label{Eq:T}\\
	\beta & < & \frac{N\Cp}{2C_0+N\Cp},\label{Eq:condition}\\
	P_{\mathrm{avg}} & = & \frac{\VCC^2}{\RL}
	\frac{2(2C_0+N\Cp)\beta}{N\Cp}\bigg/\ln\left[\frac{N\Cp+(2C_0+N\Cp)\beta}{N\Cp-(2C_0+N\Cp)\beta}\right].\label{Eq:Pavg}
\end{eqnarray}
Here, we note that these results simplify to those in~\cite{selfadaptiveWPT} when $N=1$. The results show that the WPT operation is robust against the load and coupling (receiver position) variations within the working range, similarly to~\cite{selfadaptiveWPT}.

To verify the analytical results, we have performed LTSpice simulations for different numbers of identical receivers. The system is configured with lumped elements $R_1=1~\mathrm{M\Omega}$, $R_2=3.9~\mathrm{M\Omega}$, $C_0=1~\mathrm{nF}$, $\Cp=3~\mathrm{nF}$, $\RL=100~\mathrm{kHz}$, and $V_{\mathrm{CC/EE}}=\pm 10$~V. The simulated periods for $N=2,3,...,9$ identical receivers are plotted as symbols in Fig.~\ref{Fig2}(b), and they agree very well with the analytic formula Eq.~\ref{Eq:T} (solid line). In the presentation, we will also discuss the dynamic adaptation of the system with many different receivers at variable positions.  

\section{Conclusion}
The possibility of using active and tunable metasurfaces for multiple-receiver WPT systems has been discussed. The master equations of such a WPT system have been formulated and analytical results obtained for particular examples of multiple identical receivers. The working properties of this WPT system still holds  when multiple receivers are different. However, analytical solutions  may be difficult to find as the initial conditions of $V_{3,n}$ involve time integration of all the currents. We hope that these results  will encourage future implementations of such self-adaptive smart tables for WPT applications.

\acknowledgement
This work was supported by the European Union's Horizon 2020 Future Emerging Technologies call (FETOPEN-RIA) under Grant Agreement No.~736876 (project
VISORSURF)


\end{document}